\documentclass[structabstract]{aa}
\usepackage{graphicx}
\usepackage{txfonts}
\usepackage{natbib}
\bibpunct{(}{)}{;}{a}{}{,}

\begin{document}

    \title{Galaxy morphology, luminosity, and environment in the SDSS DR7}

    \titlerunning{Environment-dependent luminosity function}

    \author{E.~Tempel\inst{1,}\inst{2} \and E.~Saar\inst{1} \and
    L.~J.~Liivam\"agi\inst{1,}\inst{2} \and A.~Tamm\inst{1} \and J.~Einasto\inst{1} \and
    M.~Einasto\inst{1} \and V.~M\"uller\inst{3}}

    \institute{Tartu Observatory, Observatooriumi~1, 61602 T\~oravere,
    Estonia \\ \email{elmo@aai.ee}
    \and Institute of Physics, Tartu University, T\"ahe~4, 51010
    Tartu, Estonia \and Astrophysikalisches Institut Potsdam, An der Sternwarte 16, 14482 Potsdam, Germany}

    \date{Received 24 November 2010 / Accepted 16 February 2011}

    \abstract
    {}
    {We study the influence of the environment on the evolution of galaxies by investigating the luminosity function (LF) of galaxies of different morphological types and colours at different environmental density levels.}  
    {We construct the LFs separately for galaxies of different morphology (spiral and elliptical) and of different colours (red and blue) using data from the Sloan Digital Sky Survey (SDSS), correcting the luminosities for the intrinsic absorption. We use the global luminosity density field to define different environments, and analyse the environmental dependence of galaxy morphology and colour. The smoothed bootstrap method is used to calculate confidence regions of the derived luminosity functions.}
    {We find a strong environmental dependency for the LF of elliptical galaxies. The LF of spiral galaxies is almost environment independent, suggesting that spiral galaxy formation mechanisms are similar in different environments. Absorption by the intrinsic dust influences the bright-end of the LF of spiral galaxies. After attenuation correction, the brightest spiral galaxies are still about 0.5\,mag less luminous than the brightest elliptical galaxies, except in the least dense environment, where spiral galaxies dominate the LF at every luminosity. Despite the extent of the SDSS survey, the influence of single rich superclusters is present in the galactic LF of the densest environment.}
    {}


    \maketitle

    \section{Introduction}

    Understanding the formation and evolution of galaxies is one of the biggest challenges of observational cosmology. The luminosity function (LF) is in this respect one of the most fundamental of all cosmological observables, helping us to describe the global properties of galaxy populations and to study the evolution of galaxies. The dependence of the LF on cosmic time, galaxy type, and environmental properties gives insight into the physical processes that govern the assembly of the stellar content of galaxies.

    The first determinations of the galaxy LF were made several decades ago \citep{Kiang:61,Christensen:75,Kirshner:79}; in the following studies, the number of galaxies used to calculate the LF has increased continuously \citep{Tully:88,Efstathiou:88,Loveday:92}. The Las Campanas Redshift Survey measured the general LF of galaxies with a good accuracy \citep{Lin:96,Bromley:98,Christlein:00}.

    Our current understanding of the general LF owes much to the 2dFGRS \citep{Norberg:02} and SDSS surveys \citep{Blanton:03,Montero-Dorta:09}. These new samples of galaxies make it possible to study the dependence of the LF on a large number of different galaxy properties, as galaxy morphology, colours, star formation rate, local and global density environment, etc.

    The morphology of a given galaxy is a reflection of its merger history. In studies of the LF, galaxy morphology has been determined either by its colours \citep{Yang:09}, spectra \citep{Folkes:99,Madgwick:02,de-Lapparent:03} or photometric profile \citep{Bell:03,Driver:07a}; some studies use artificial neural networks for morphological classification \citep{Ball:06}. The most accurate, but by far the most time consuming approach is to use visual classification \citep{Marzke:94a,Marzke:98,Kochanek:01,Cuesta-Bolao:03,Nakamura:03}. For the SDSS survey, visual classification has become possible thanks to the Galaxy Zoo project \citep{Lintott:08} that will help us to study the morphology and the LF in detail in the future. Recently \citet{Huertas-Company:11} have published an automated morphological classification based on the Galaxy Zoo data. In all these studies, the classification of early-type and late-type galaxies is different, but all studies agree that later-type galaxies have a fainter characteristic magnitude and a steeper faint-end slope of the LF. The biggest differences in previous studies are found at the faint-end of the LF, where classification is less certain than for brighter galaxies.

    To understand how galaxies form, we also need to understand where galaxies are located; it is essential to study the LF dependence on the environment. It is well known from the halo occupation distribution models that the local environment is crucial for the galaxy distribution \citep[e.g.][]{Zandivarez:06,Park:07}: luminous galaxies tend to occupy high mass haloes and low luminosity galaxies reside mainly in low mass haloes. This motivates the study of the LF in galaxy groups \citep{Xia:06,Zandivarez:06,Hansen:09,Yang:09}.

    A likewise important, but not so well understood factor is the global environment where the galaxy is located -- its place in the supercluster-void network. In \citet{Tempel:09} we have found that the global environment has an important role in determining galaxy properties. Some studies have been dedicated only to special regions: e.g., \citet{Mercurio:06} investigate the Shapley supercluster. The dependence on the environment has been also studied numerically \citep{Mo:04} and using semi-analytical models \citep{Benson:03a,Khochfar:07}. These semi-analytical models allow us to study morphological evolution: how the morphology of a galaxy changes in time. To compare these models with the real Universe, we need to know the observed LF in detail.

    The influence of the global environment on the LF has been investigated by \citet{Hoyle:05}, using the SDSS data, and by \citet{Croton:05}, using the 2dFGRS data. These results show strong environmental trends: galaxies in higher density regions tend to be redder, of earlier type, have a lower star formation rate, and are more strongly clustered. Some of these trends can be explained with the well known morphology-density relation \citep{Einasto:74,Dressler:80,Postman:84} and luminosity-density relation \citep{Hamilton:88}. It is less well known how far these trends extend when moving toward extreme environments, into deep voids or superclusters.

   Recent studies have shown that dust plays an important role in galaxy evolution and it may significantly influence the luminosities and colours of galaxies \citep{Pierini:04,Tuffs:04,Driver:07,Rocha:08,Tempel:10}, especially for late-type galaxies. Thus, in order to study intrinsic properties of galaxies, it is necessary to take dust extinction into account. Using the SDSS data, \citet{Shao:07} have studied the influence of dust on the LF. In general, dust is important for late-type spiral galaxies; nearly edge-on galaxies are most affected.

    In the present paper we use the SDSS data to study the LF in different global environments and for different types of galaxies, taking the effect of dust attenuation into account. The LF dependency on group properties will be analysed in a forthcoming paper.

    Throughout this paper we assume a Friedmann-Robertson-Walker cosmological model with the total matter density $\Omega_\mathrm{m}=0.27$, dark energy density $\Omega_\Lambda=0.73$, and the Hubble constant $H_0=100\,h\,\mathrm{km\,s^{-1}Mpc^{-1}}$. Magnitudes are quoted in the AB~system.

    \section{Data and method of analysis} 

    \subsection{Galaxy sample} 

    Our present analysis is based on the SDSS Data Release~7 \citep{Abazajian:09}. We used the main galaxy sample, downloaded from the Data Archive Server (DAS) of the SDSS. Additionally, some of the galaxy parameters that were not available from the DAS, were downloaded from the Catalog Archive Server (CAS) of the SDSS using a Structured Query Language (SQL) search. We use only the contiguous imaging and spectroscopic area of the Northern Galactic Cap (the Legacy Survey), which covers 7646 square degrees of the sky. The main galaxy sample of the SDSS includes galaxies brighter than the limiting $r$-band Petrosian magnitude 17.77. Our galaxy sample is described in detail in \citet{Tago:10}; in total, this sample includes 583362 galaxies.

    For the present work we use galaxies that lie in the distance interval 55--565\,$h^{-1}$Mpc. We use the co-moving distances \citep[see, e.g.][]{Martinez:02} as the linear dimensions, calculated using the cosmological parameters listed above. The lower distance limit (55\,$h^{-1}$Mpc) is chosen to exclude galaxies of the Local Supercluster; the upper limit (565\,$h^{-1}$Mpc) has been set because the SDSS sample becomes very diluted at large distances. Since the faint limiting magnitude has varied throughout the survey (between 17.5 and 17.77), we shall use the limiting magnitude 17.6 in the $r$-band; this leads to a uniform distribution for the $V/V_\mathrm{max}$ (see Sect.~\ref{sec:lumfun}). In addition to the lower limiting magnitude 17.6, we use the upper limiting $r$-band magnitude 14.5, since the brighter end of the SDSS sample is incomplete due to saturation of bright galaxy images and due to blending with bright saturated stars. As suggested in the SDSS webpage\footnote{http://www.sdss.org/dr7/algorithms/photometry.html\#which\_mags}, we use composite model magnitudes\footnote{http://www.sdss.org/dr7/algorithms/photometry.html\#cmodel}.

    \begin{table}
        \caption{Galaxy parameters used.}
        \label{tab:param}
        \centering
        \begin{tabular}{lll}
            \hline\hline
            Name\tablefootmark{a} & SDSS Name\tablefootmark{b} & Description \\
            \hline
            $m_{x}$\tablefootmark{c} & -- & extinction-corrected\\
            &  & \emph{cmodel} magnitude\\
            $M_x$ & -- & absolute magnitude \\
            $z$ & z & spectroscopic redshift \\
            $d$ & -- & co-moving distance \\
            $f_{\mathrm{deV}}$ & fracDeV\_r\tablefootmark{d} & weight of the $r^{1/4}$ component \\
            $q_{\mathrm{exp}}$ & expAB\_r & exponential fit $a/b$ \\
            $q_{\mathrm{deV}}$ & deVAB\_r & de~Vaucouleurs fit $a/b$ \\
            $r_{\mathrm{exp}}$ & expRad\_r & exponential fit scale radius \\
            \hline
        \end{tabular}
        \tablefoot{
            \tablefoottext{a}{Parameter name, as used in the present paper.} \tablefoottext{b}{Parameter name as given in the SDSS CAS archive.} \tablefoottext{c}{$x$ is either $u$, $g$, $r$, $i$ or $z$ filter in the SDSS; the magnitude $m_x$ is corrected for the Galactic extinction.} \tablefoottext{d}{Weight of the de~Vaucouleurs component in the best-fit composite model.}
        }
    \end{table}

    All the galaxy parameters used are given in Table~\ref{tab:param} together with their SDSS CAS archive names. The apparent magnitude $m$ was transferred into the absolute magnitude $M$ according to the usual formula
    \begin{equation}\label{eq:absmag}
        M_\lambda=m_\lambda-25-5\log_{10}(d_L)-K,
    \end{equation}
    where the luminosity distance $d_L=d(1+z)$; $d$ is the co-moving distance in the units of $h^{-1}\mathrm{Mpc}$ and $z$ is the observed redshift. The term $K$ is the $k$+$e$-correction. The apparent magnitude $m_\lambda$ is the composite model flux, calculated according to the SDSS webpage\footnotemark[2], corrected for Galactic extinction according to \citet{Schlegel:98}.

    The $k$-corrections for the SDSS galaxies were calculated using the \mbox{KCORRECT} algorithm (version \mbox{v4\_1\_4}) developed by \citet{Blanton:03a} and \citet{Blanton:07}. The evolution corrections $e$ have been applied according to \citet{Blanton:03}.

    \subsection{Selection effects and luminosity function estimation} 
    \label{sec:lumfun}
    
    Several methods have been used in the past to estimate the luminosity functions: $V_{\mathrm{max}}^{-1}$, $C^-$, various maximum-likelihood estimators \citep[see, e.g.][]{willmer:97,wall:03}. The main reasons for so many methods were small sample volumes and, consequently, small numbers of galaxies. For such samples, deviations from overall homogeneity influenced the results, especially for the simplest, $V_{\mathrm{max}}^{-1}$ method.
    
    Nowadays, both the sample volumes and sample sizes are about two orders of magnitude larger than in the recent past, and all methods work equally well. We chose the $V_{\mathrm{max}}^{-1}$ method as the physically most transparent and statistically straightforward. Since the luminosity function is practically a probability distribution, it should be estimated as a distribution, and this is what the $V_{\mathrm{max}}^{-1}$ and the $C^-$ methods do. The $C^-$ method gives the integral distribution, thus its errors are more difficult to estimate than for the $V_{\mathrm{max}}^{-1}$ method. Estimating probability distributions is a well-developed topic in statistics, and we can choose all the modern tools -- kernel densities, adaptive kernels, smoothed bootstrap for pointwise confidence intervals (see Appendix~\ref{app:1}).
    
    The only requirement for the $V_{\mathrm{max}}^{-1}$ method is that, on an average, the spatial distribution of galaxies should be roughly uniform. That can be easily tested; we did that.
    
    Below we give a short summary of the $V_{\mathrm{max}}^{-1}$ method; a  detailed description with our modifications is given in Appendix~\ref{app:1}.
    
    \begin{figure}
        \includegraphics[width=8.8cm]{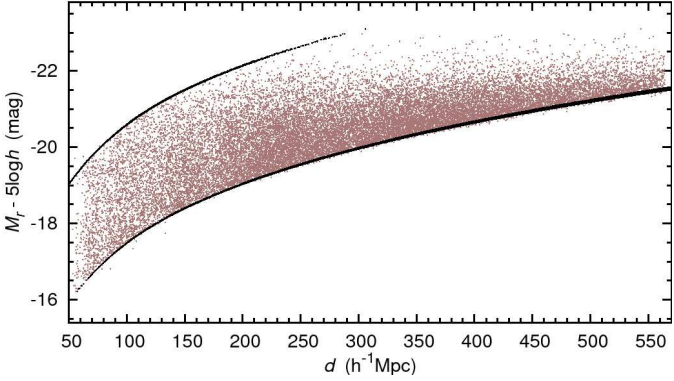}
        \caption{The absolute magnitude of the SDSS DR7 main sample in the $r$-band versus distance. Black solid lines show the upper and lower limits of the flux-limited sample: 14.5 and 17.6\,mag in the observer frame $r$-band, respectively, with the average $K$-corrections.}
        \label{fig:dist_vs_mag}
    \end{figure}

    \begin{figure}
        \includegraphics[width=8.8cm]{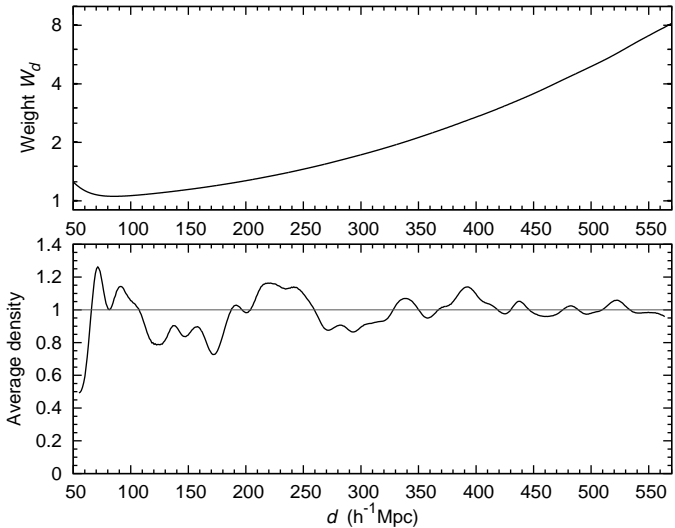}
        \caption{\emph{Upper panel}: the weight factor $W_d$ at different distances. \emph{Lower panel}: the average global density in thin concentric shells and given in units of mean density (calculated using the weight factor $W_d$).}
        \label{fig:weight}
    \end{figure}

    In a flux-limited sample, the principal selection effect is the absence of galaxies fainter than the survey limiting magnitude. This effect is well seen in Fig.~\ref{fig:dist_vs_mag}, showing the absolute luminosities of galaxies plotted against their distance; at large distances only the brightest galaxies are seen.

    To take this effect into account in the determination of the LF of galaxies we used the standard $V_{\mathrm{max}}^{-1}$ weighting procedure. The differential luminosity function $n(L)\mathrm{d}L$ (the expectation of the number density of galaxies of the luminosity $L$) was found as usual:
    \begin{equation}
        n(L) {\mathrm d}L = \sum_i\frac{\mathbf{I}_{(L,L+\mathrm{d}L)}(L_i)}
        {V_{\mathrm{max}}(L_i)},
        \label{eq:lf}
    \end{equation}
    where $\mathrm{d}L$ is the luminosity bin width, $\mathbf{I}_A(x)$ is an indicator function, selecting galaxies that belong to a particular luminosity bin; $V_{\mathrm{max}}(L)$ is the maximum volume where a galaxy of a luminosity $L$ can be observed in the present survey; summation is made over all galaxies of the survey. This procedure is non-parametric, and gives both the form and the true normalisation of the LF.
    
As said above, the spatial distribution of galaxies in the sample should be uniform. This can be tested by examining the distribution of $V/V_{\mathrm{max}}$, where $V=V(d)$ is the part of the sample volume closer than the distance $d$ to a galaxy \citep[see, e.g.][]{wall:03} -- this distribution should be uniform. Due to the large size of our sample it is indeed practically uniform -- the \mbox{K-S} test gives the probability (1--2)$\times10^{-7}$ for the distributions to differ. The $V/V_{\mathrm{max}}$ distribution is sensitive to sample incompleteness; choosing the limiting $r$-band magnitude 17.6 instead of the standard 17.77 was essential to make it practically uniform.

For a further test, we generated several test samples of 500000 galaxies within the same distance and magnitude limits as for the observational sample, using the Schechter luminosity function with typical parameters ($\alpha=-1.14, M^{\star}=-20.6$), and our method recovered the initial parameters with deviations of 0.03\% ($\sigma_{\alpha}=0.0003, \sigma_{M^\star}=0.003$).

    As the luminosity function is rapidly changing with luminosity, especially at the bright end, varying bin widths should be used. This is most easily achieved by an adaptive kernel estimation of the LF; the method is explained in the Appendix~\ref{app:1}. Uncertainties of the LF (the confidence regions of the estimate) were estimated by the smoothed bootstrap method, also described in the Appendix~\ref{app:1}.

    The maximum volume $V_{\mathrm{max}}(L)$, where a galaxy of an absolute luminosity $L$ could be observed is set by the minimum and maximum distance at which the corresponding apparent luminosity would fall within the luminosity limits. The distance limits are shown as solid lines in Fig.~\ref{fig:dist_vs_mag}. These limits were calculated, using Eq.~(\ref{eq:absmag}), where the apparent magnitude $m$ was set to the survey limiting magnitudes: 14.5 and 17.6 in $r$-band. Since the $k$-correction is colour- and distance-dependent, we used the colour-averaged values of the $k$-correction for survey galaxies found as a function of distance.

    \subsection{Determining environmental densities} 

    The environmental effects on the LF were studied at the level of the global density field, the so called supercluster-void network. We use the luminosity density for calculating the global density, assuming that it is proportional to the total matter density.

    In calculations of the luminosity density, selection effects typical to flux-limited samples had to be considered: in distant fields, only the brightest galaxies have been detected, while in nearby regions, the brightest galaxies are absent (see Fig.~\ref{fig:dist_vs_mag}). To take this effect into account, we calculated the distance-dependent weight factor $W_d$
    \begin{equation}
        W_d =  {\frac{\int_0^\infty L\,n
        (L)\mathrm{d}L}{\int_{L_1}^{L_2} L\,n(L)\mathrm{d}L}} ,
        \label{eq:weight}
    \end{equation}
    where $L_{1,2}=L_{\sun} 10^{0.4(M_{\sun}-M_{1,2})}$ are the luminosity limits of the observational window at a distance $d$, corresponding to the absolute magnitude limits of the window $M_1$ and $M_2$; we took $M_{\sun}=4.64$\,mag in the $r$-band \citep{Blanton:07}. The limits $M_1$ and $M_2$ have been chosen the same as for the LF (black lines in Fig.~\ref{fig:dist_vs_mag}). In the latter equation, $n(L)$ is taken to be the LF in the $r$-band for all galaxies (see Table~\ref{table:lf}), because we are interested in the total luminosity density field.
    
    We assumed that every galaxy belongs to a group of galaxies, several of which may lie outside the observational window. The luminosity of the unobserved galaxies was taken into account by multiplying the observed galaxy luminosities by the weight $W_d$. Due to their peculiar velocities, the distances of galaxies are somewhat uncertain; if the galaxy belongs to a group, we use the group distance to determine the weight factor. For the luminosity density field calculations we used the luminosities without the intrinsic absorption corrections.

    The luminosity function (calculated as described in the Appendix~\ref{app:1}) was approximated by a double-power law:
    \begin{equation}
        n (L) \mathrm{d}L \propto (L/L^{*})^\alpha (1 +
        (L/L^{*})^\gamma)^{(\delta-\alpha)/\gamma}
        \mathrm{d}(L/L^{*}),
        \label{eq:lf_dpl}
    \end{equation}
    where $\alpha$ is the exponent at low luminosities $(L/L^{*}) \ll 1$, $\delta$ is the exponent at high luminosities $(L/L^{*}) \gg 1$, $\gamma$ is a parameter that determines the speed of the transition between the two power laws, and $L^{*}$ is the characteristic luminosity of the transition. A similar double-power law was also used by \citet{Vale:04} to fit the mass-luminosity relation in their subhalo model and by \citet{Cooray:05} to fit the LF of central galaxies. Since several papers have shown that the Schechter function is not a best fit for the LF \citep{Blanton:05,Mercurio:06,Yang:08a,Tempel:09}, especially at the bright end, we shall use the double-power law for analytical approximation. Parameters for the LF approximations are given in Table~\ref{table:lf}. We give there also the fits for the Schechter function, to facilitate comparison with the results of other authors.

    We used the $B_3$ spline kernel to calculate the density field. We chose a 8\,$h^{-1}$Mpc scale smoothing for global densities \citep{Einasto:07b}. Details of the density calculation are given in \citet{Liivamagi:11}.

    The upper panel of Fig.~\ref{fig:weight} shows the weight factor $W_d$ as a function of distance; the lower panel of Fig.~\ref{fig:weight} shows the average global density at different distances. When the weight factor is applied, the global density becomes roughly constant; the remaining variations are due to the large-scale structure.

    We split our sample into four global density regions: voids, superclusters, and two intermediate regions. Both the void and the supercluster regions contain approximately 10\% of all galaxies. The rest of the galaxies are divided nearly equally between the two other regions. We designate these four regions as D1, D2, D3, and D4, where D1 is the void region and D4 is the supercluster region; D2 and D3 are the two intermediate density regions. The number of galaxies in each density region and the corresponding volume are given in Table~\ref{tab:env}.
    \begin{table}
        \caption{Properties of different environmental regions.}
        \label{tab:env}
        \centering
        \begin{tabular}{llrrrr}
            \hline\hline
        Region & Densities\tablefootmark{a} & $N_\mathrm{gal}$ & $N_\mathrm{spiral}$ & $N_\mathrm{ellipt.}$ & V\tablefootmark{b} \\
            \hline
            D1 & 0.0--0.8 & 54942 & 32963 &  7836 & 57.6  \\
            D2 & 0.8--2.0 & 173320 & 86239 & 36718 & 29.1  \\
            D3 & 2.0--5.0 & 208210 & 85749 & 57965 &  12.0 \\
            D4 & 5.0--$\infty$ & 66884 & 23123 & 22026 & 1.3  \\
            \hline
        \end{tabular}
        \tablefoot{
            \tablefoottext{a}{Densities are given in units of the global mean density ($0.01526\times10^{10}\,h\mathrm{L}_{\sun}\mathrm{Mpc}^{-3}$).} \tablefoottext{b}{Percent of the total volume.}
        }
    \end{table}


    \section{Dust attenuation in galaxies} 

    In this section we describe the necessary steps to correct the galaxy luminosities for internal attenuation. Since dust affects mostly late-type, spiral galaxies, we start this section with classifying galaxies into spirals and ellipticals. Since dust attenuation also depends on the galaxy inclination angle and galaxy colour (or galaxy type), we take the necessary steps to take that into account. We will end this section by showing how our attenuation correction works.

    \subsection{Galaxy classification}\label{sect:morf} 

    Spiral galaxies have more dust than ellipticals and therefore the observed luminosities of spiral galaxies are more affected by dust. If we want to correct for dust attenuation, we have to know the morphology of a galaxy. Since dust affects mainly spiral galaxies, we will compose a sample of spiral and/or disc-dominated galaxies. We will use this sample to study the effect of dust attenuation. Additionally to spiral galaxies, we will compose also a sample of elliptical and/or bulge dominated S0 galaxies. We will use this elliptical galaxy sample only for comparison, and not for a detailed study.

    As one source, we will use the morphological classification by the Galaxy Zoo project \citep{Lintott:08}. The Galaxy Zoo project has led to the morphological classification of nearly one million objects from the SDSS data by visual inspection. \citet{Banerji:10} used the Galaxy Zoo data to develop a machine learning algorithm (an artificial neural network) for galaxy classification; they have published also distribution histograms for different parameters of various types of objects: stars, spirals, ellipticals, and mergers. In this paper we use these histograms as the basis for selecting dominantly spiral or dominantly elliptical galaxies. Additionally, we will use the galaxy colour distributions from \citet{Lintott:08}.

    \begin{figure}
        \includegraphics[width=8.8cm]{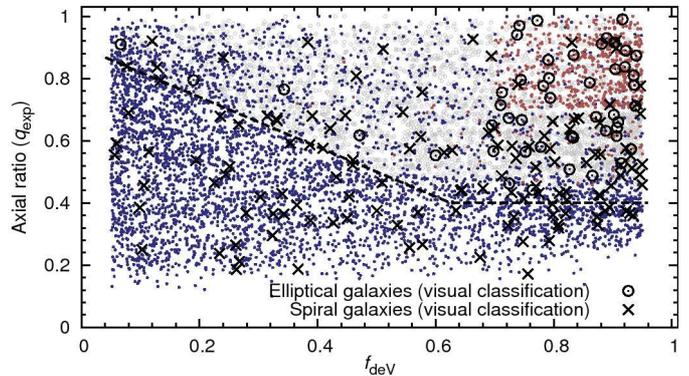}
        \caption{Classification of galaxies into spirals and ellipticals. Red points (mostly in the upper right corner) are ellipticals and blue points (mostly below the dashed line) are spirals. Empty circles are elliptical galaxies classified by visual inspection; crosses are visually classified spiral galaxies. Grey points are galaxies where classification is unclear. The dashed line shows our spiral galaxy selection criterion (see text).}
        \label{fig:classific}
    \end{figure}

    \begin{figure}
        \includegraphics[width=8.8cm]{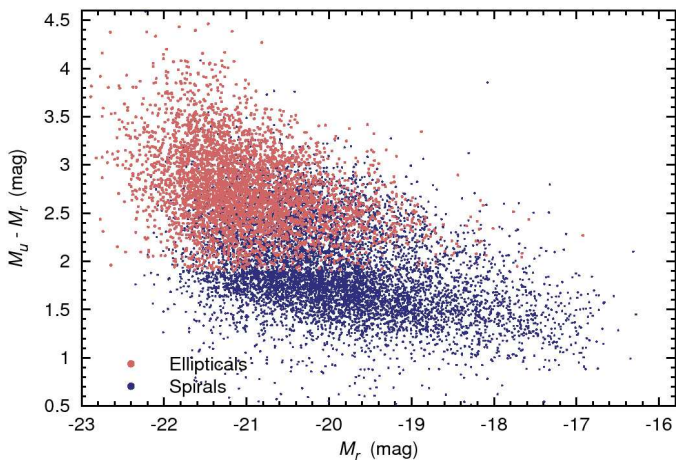}
        \caption{Colour-magnitude diagram for spiral and elliptical galaxies. Spirals are marked by blue dots (the lower cloud) and ellipticals by red dots (the upper cloud).}
        \label{fig:colmag}
    \end{figure}

    We will compare the distributions from the Galaxy Zoo project with our own visual classifications. We have classified a small sample (of nearly one thousand) galaxies in the Sloan Great Wall region \citep{Einasto:10}. In this paper we use only galaxies with a clear classification to test the distributions taken from the Galaxy Zoo project. Additionally, we found another criterion ($f_{\mathrm{deV}}$) to classify galaxies into spirals and ellipticals.

    In the SDSS one of the main parameters that determines the type of a galaxy is the photometric parameter $f_{\mathrm{deV}}$ (see Table~\ref{tab:param}) -- the point-spread function corrected indicator of galaxy morphology. The surface luminosity distribution of each galaxy in the SDSS has been fitted by the exponential and the de~Vaucouleurs profiles. The best linear combination of these is used to represent the profile of the galaxy, and $f_{\mathrm{deV}}$ indicates the fraction of luminosity contributed by the de~Vaucouleurs profile. \citet{Bernardi:05} used $f_{\mathrm{deV}} > 0.8$ to select early-type galaxies. \citet{Shao:07} used $f_{\mathrm{deV}} < 0.5$ to select galaxies which are dominated by the exponential component (i.e. spiral galaxies).
    
    \begin{figure}
        \includegraphics[width=8.8cm]{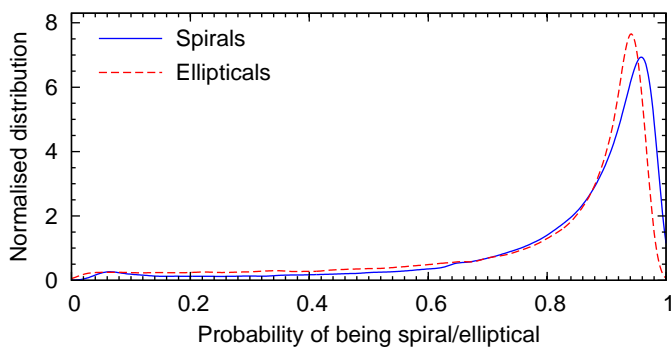}
        \caption{Morphological type probability distributions according to \citet{Huertas-Company:11} of galaxies classified as spiral (blue solid line) and ellipticals (red dashed line) in this paper. In our final sample, only galaxies with the given probability at least 0.5 were included.}
        \label{fig:huertas}
    \end{figure}

    We combine the value $f_{\mathrm{deV}}$ with the exponential profile axis ratio $q_{\mathrm{exp}}$ to do the primary classification. In Fig.~\ref{fig:classific} we show the $f_{\mathrm{deV}}$ versus $q_{\mathrm{exp}}$ plot; crosses are spiral galaxies and circles are elliptical galaxies as classified by \citet{Einasto:10}. As seen in this Figure, all galaxies, which have $q_{\mathrm{exp}} < 0.4$, are spirals; the corresponding distribution in \citet{Banerji:10} confirms that. \citet{van-den-Bosch:09} have also shown that for elliptical galaxies, the axial ratio is mostly greater than 0.5. We also see that when moving toward lower values of $f_{\mathrm{deV}}$, the spiral dominated region becomes larger: the value of $q_{\mathrm{exp}}$ can be larger. This is expected, since low values of $f_{\mathrm{deV}}$ point to disc-dominated objects. In our classification we will take this behaviour into account. The dashed line in Fig.~\ref{fig:classific} shows the limit for spiral galaxies. We will also leave unclassified those galaxies where $f_{\mathrm{deV}}>0.95$ or $f_{\mathrm{deV}}<0.05$, since classification at these extreme values may be rather uncertain.

    For all other galaxies that cannot be classified directly in the $f_{\mathrm{deV}}$ versus $q_{\mathrm{exp}}$ plot, we use the galaxy colour as an additional parameter. Distributions of galaxy colours for different types of galaxies have been taken from \citet{Lintott:08} and \citet{Banerji:10}; additionally, we have checked these distributions, using our own small sample of visually classified galaxies. Rather tight (and thus secure) constraints have been chosen for the colour criteria, since each criterion is used to add galaxies to the sample. If a galaxy satisfied one criterion, then the following criteria were ignored. In the brackets we give the cumulative numbers of galaxies in the spiral/elliptical samples, after applying this criterion. The criteria are:
    \begin{itemize}
        \item all galaxies, where $q_{\mathrm{exp}}<0.4$ or $q_{\mathrm{deV}}<0.4$ (and $0.05\le f_{\mathrm{deV}}\le 0.95$), are classified as spirals \citep{Banerji:10} (60335/0),
        \item galaxies, where $q_{\mathrm{exp}}<0.9-0.8f_{\mathrm{deV}}$ (and $0.05\le f_{\mathrm{deV}}\le 0.95$), are spirals (Fig.~\ref{fig:classific}) (105819/0),
        \item galaxies, where the rest-frame colour $M_u$-$M_r>3.0$, are ellipticals \citep[][and Fig.~\ref{fig:colmag}]{Lintott:08} (105819/40389),
        \item galaxies, where the rest-frame colour $M_u$-$M_r<1.9$, are spirals \citep{Lintott:08} (208483/40389),
        \item galaxies, where the observer-frame $m_g$-$m_r>1.2$, are ellipticals \citep{Banerji:10} (208483/46237),
        \item galaxies, where the observer-frame $m_g$-$m_r<0.6$, are spirals \citep{Banerji:10} (216822/46237),
        \item galaxies, where $q_{\mathrm{exp}}<0.45$ or $q_{\mathrm{deV}}<0.45$, are classified as spirals \citep{van-den-Bosch:09,Banerji:10} (248604/46237),
        \item galaxies, where $f_{\mathrm{deV}}>0.7$ and $q_{\mathrm{exp}}>0.7$, are ellipticals (Fig.~\ref{fig:classific}) (248604/142023).
    \end{itemize}
    The last criterion is included to add mostly ellipticals and/or bulge dominated S0 galaxies into our elliptical sample. All the other galaxies, not classified according to the criteria given above, are used only to study the LF of the total galaxy sample. For the final refinement of the sample we exploited the automatic classification recently published by \citet{Huertas-Company:11}. In this paper, the authors assign for every galaxy a probability of being early- or late-type. Figure~\ref{fig:huertas} shows the distribution of probabilities of being early- or late-type galaxies for our classified spirals and ellipticals. It is seen that our classification agrees very well with the classification by \citet{Huertas-Company:11}. However, to further increase the reliability of our classification, we excluded from our sample all the galaxies, where this probability was less than 0.5, leaving us with the final sample of 228074 spirals and 124545 ellipticals.

    Figure~\ref{fig:classific} shows our classification in the $f_{\mathrm{deV}}$ versus $q_{\mathrm{exp}}$ plot: all galaxies, which are marked by blue points, are classified as spirals; all galaxies, which are marked with red points, are classified as ellipticals; grey points are non-classified galaxies. Figure~\ref{fig:colmag} shows the classical colour-magnitude diagram for our galaxies; spirals are statistically bluer and fainter, and ellipticals are brighter and redder. Since we use also other parameters for galaxy classification, some of the spirals are located in the region, where mostly ellipticals reside: however, the number of these spirals ($M_u-M_r>2.3$) is relatively small (20\% of all spirals). Most of these red spirals are those where $f_{\mathrm{deV}}$ is large: e.g. their luminosity profile is bulge dominated, but the visible axial ratio ($q_{\mathrm{exp}}$) is small and therefore we have classified these galaxies as spirals. Practically all of the disc-dominated spirals are located below elliptical galaxies in this colour-magnitude diagram. Using the Galaxy Zoo data, \citet{Skibba:09} found that red spirals account for nearly a quarter of all the spirals, which is in accordance with our findings.

    \begin{figure}
        \includegraphics[width=8.8cm]{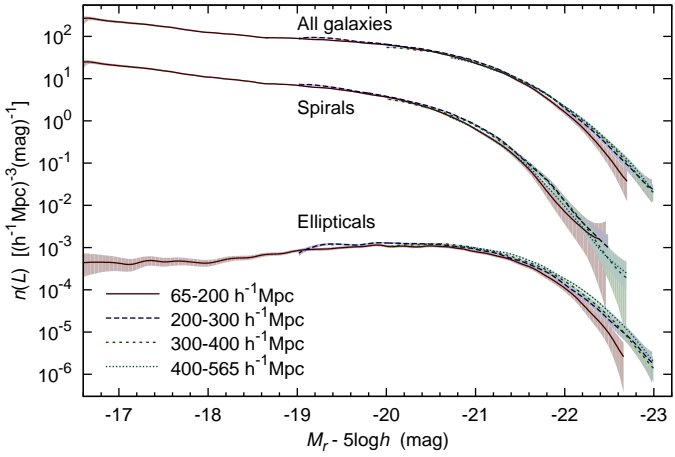}
        \caption{LF for different distance intervals. LF of spirals and all galaxies are shifted up by three and four units in the logarithmic scale, respectively. The filled areas show the 95\% confidence regions of the LF.}
        \label{fig:lf_dist_type}
    \end{figure}

    In our classification about half of the galaxies (45\%) are spirals, about one quarter (25\%) are ellipticals and for 30\% of the galaxies, the classification is unclear.

    To test the reliability of our classification, we calculated the luminosity functions for different distance intervals: 65--200, 200--300, 300--400, and 400--565\,$h^{-1}$Mpc. The LFs for different types of galaxies are shown in Fig.~\ref{fig:lf_dist_type}. This figure shows that the classification is statistically correct since the LF for all cases is nearly distance independent.

    \begin{figure}
        \includegraphics[width=8.8cm]{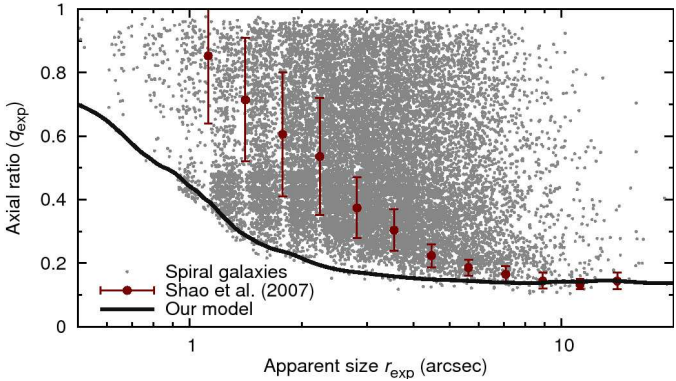}
        \caption{Apparent axial ratio versus apparent size for spiral galaxies. Solid line is our fit for edge-on spiral galaxies. Red points with errorbars are taken from \citet{Shao:07}.}
        \label{fig:rad_vs_ab}
    \end{figure}

    \subsection{Restoring the intrinsic inclination angle} 

    The intrinsic absorption is closely related to the morphology of galaxies. Additionally, attenuation depends on the inclination angle \citep{Tuffs:04,Driver:07,Tempel:10}. To take this effect into account, we need to know the intrinsic inclination for every galaxy. Since the intrinsic inclination angle depends both on the visible and the intrinsic axial ratios, we can restore the intrinsic inclination angle only statistically. In order to consider dust attenuation, the inclination angle is needed only for spiral galaxies, but we will calculate it also for elliptical galaxies.

    \begin{figure}
        \includegraphics[width=8.8cm]{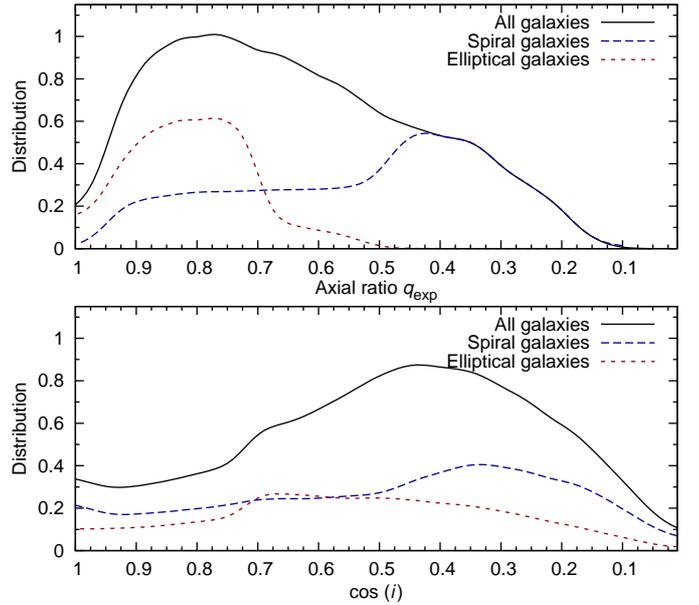}
        \caption{\emph{Upper panel}: distribution of the apparent axis ratio for all, for spiral, and for elliptical galaxies. \emph{Lower panel}: distribution of the restored intrinsic inclination angle for all, for spiral, and for elliptical galaxies.}
        \label{fig:inc_stat}
    \end{figure}

    Assuming that we know the axial ratios and the intrinsic inclination angle of a galaxy, we can calculate the apparent axial ratio $q\equiv b/a$ using the expression given by \citet{Binney:85}
    \begin{equation}
        \left(\frac{b}{a}\right)^2=\frac{A+C-\sqrt{(A-C)^2+B^2}}{A+C+\sqrt{(A-C)^2+B^2}},
        \label{eq:binney_valem}
    \end{equation}
    \begin{eqnarray}
        A&\equiv& \frac{\cos^2{\theta}}{\xi^2}\left(\sin^2{\phi}+
        \frac{\cos^2{\phi}}{\zeta^2}\right) + \frac{\sin^2{\theta}}{\zeta^2} ,\\
        B&\equiv& \left(1-\frac{1}{\zeta^2}\right)\frac{1}{\xi^2}\cos\theta\sin{2\phi} ,\\
        C&\equiv& \left(\frac{\sin^2{\phi}}{\zeta^2}+\cos^2{\phi}\right)\frac{1}{\xi^2} .
    \end{eqnarray}
    In the last equation, $1 \ge \zeta \ge \xi$, $\xi$ is the ratio of the shortest semi-axis to the longest semi-axis and $\zeta$ is the ratio of the two longer semi-axes. $\theta$ is the inclination of a galaxy -- the angle between the plane of the galaxy and the plane of the sky; $\phi$ is the angle between the longest semi-axis and the line of sight.

    The SDSS gives the apparent axial ratio $q$ of each galaxy image. For spiral galaxies we use the $r$-band axial ratio $q_{\mathrm{exp}}$, taken from the best fit of the image with an exponential profile convolved with the point spread function \citep{Stoughton:02}.

    To calculate the inclination angle $\theta$ from Eq.~(\ref{eq:binney_valem}), we use a statistical approach. We assume that the longest semi-axes are randomly oriented in space and the ratios $\xi$ and $\zeta$ are random, with different probability distributions for spirals and ellipticals. We use the Monte Carlo method to select random values for inclination angle and axial ratios.

    To find the intrinsic ratio of the shortest semi-axis to the longest semi-axis, we use galaxies with a small value of $q_{\mathrm{exp}}$ and assume that this value is the ratio $\xi$. As demonstrated by \citet{Shao:07}, the apparent axial ratio depends on the apparent size and therefore, the value of $\xi$ depends also on it. To find how the intrinsic thickness of an observed spiral galaxy depends on its apparent size, we plot the apparent axial ratio versus the apparent size (see Fig.~\ref{fig:rad_vs_ab}) and for every apparent size, we find the minimum thickness as shown with solid line in Fig.~\ref{fig:rad_vs_ab}. For every spiral galaxy, we find the ratio $\xi$ according to the scaling in Fig.~\ref{fig:rad_vs_ab} (solid line). We selected edge-on galaxies for deriving the scaling, thus the ratio $\xi$ should be statistically correct. In the same figure, results from \citet{Shao:07} are shown; our distribution remains somewhat lower than that of \citet{Shao:07}. The difference arises from different methods for selecting spiral galaxies. \citet{Shao:07} have used only $f_{\mathrm{deV}}<0.5$ to select spiral galaxies; this criterion alone is not reliable for smaller galaxies. There are many galaxies, where $f_{\mathrm{deV}}>0.5$ but $q_{\mathrm{exp}}<0.5$ -- in our classification, these galaxies are spirals and are located below the \citet{Shao:07} region in Fig.~\ref{fig:rad_vs_ab}.

    For elliptical galaxies the ratio of the shortest semi-axis to the longest semi-axis is taken to be 0.7, with an 1-$\sigma$ error 0.1. This is in agreement with \cite{van-den-Bosch:09}, who derived the axial ratios for 13 elliptical galaxies. Additionally, the ratio $\xi$ must be smaller than the apparent axial ratio by definition.

    Since the discs of spiral galaxies are not round and elliptical galaxies are triaxial, the ratio $\zeta$ is less than one. \citet{Ryden:04} showed that the intrinsic axial ratio of discs is $0.85^{+0.1}_{-0.2}$, which is slightly smaller than derived by \citet{Andersen:02} ($0.9^{+0.06}_{-0.18}$). Both values are close to that derived for elliptical galaxies by \citet{van-den-Bosch:09}. In this paper we adopt the $\zeta$ distribution with the maximum at 0.9; toward larger values, the 2-$\sigma$ deviation is 0.05, and toward smaller values this deviation is 0.1. Additionally we demand that the value of $\zeta$ has to be larger than the shortest-to-longest axis ratio. We will use the same $\zeta$ distribution both for spiral and elliptical galaxies.

    Figure~\ref{fig:inc_stat} shows the distribution of $q$ for all galaxies, for spiral, and for elliptical galaxies separately. If the ratio $q$ were caused only by the inclination of a galaxy, then, assuming that the rotation axes were randomly oriented, the probability distribution of $q$ should be flat. As we see, the distribution is not flat and decreases toward both smaller and larger axial ratios.

    The results of restoring the inclination angles are presented in the lower panel of Fig.~\ref{fig:inc_stat}. The distribution of inclination angles is not perfectly random, but is much closer to that, than the apparent axial ratio distribution, especially for elliptical and spirals separately.

    \subsection{Accounting for dust attenuation in spiral galaxies} 

    \begin{figure}
        \includegraphics[width=8.8cm]{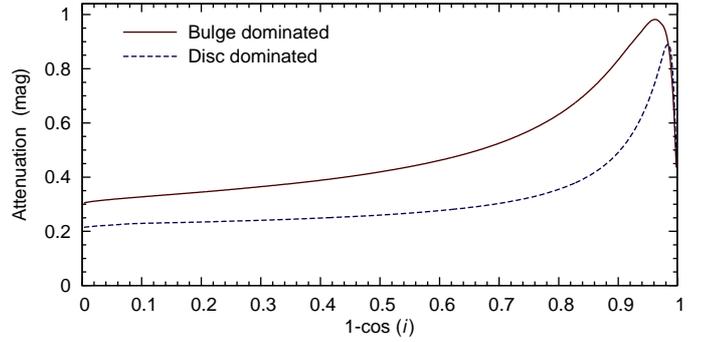}
        \caption{The dependence of attenuation on galaxy inclination for disc-dominated and bulge-dominated spiral galaxies.}
        \label{fig:nld_model}
    \end{figure}
    
    \begin{figure}
        \includegraphics[width=8.8cm]{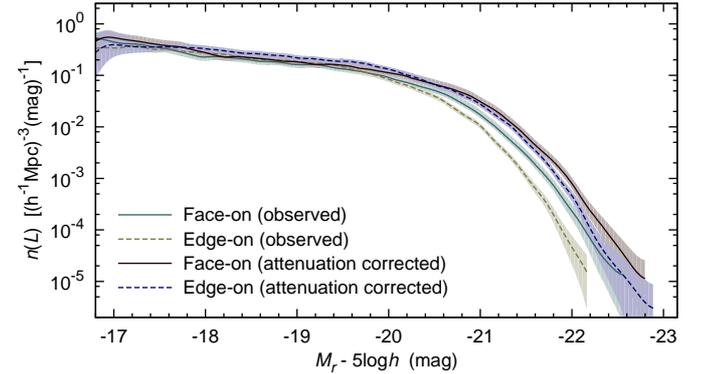}
        \caption{Luminosity functions for the edge-on and face-on samples of spiral galaxies. Darker lines are attenuation-corrected LFs, dimmer lines are the LF for observed luminosities. Filled areas show the 95\% confidence intervals.}
        \label{fig:lf_incl}
    \end{figure}

    In this section we describe the necessary steps to take dust attenuation into account. We will use the morphological classification, derived in Sect.~\ref{sect:morf}; dust attenuation will be considered for spiral galaxies only. Late-type spiral galaxies have generally more dust than early-type spirals, thus dust attenuation is higher there; we will take that additionally into account. It is known that blue galaxies have more dust and therefore the attenuation is larger; for redder galaxies dust attenuation is less important. The galaxy colour is also an indicator of the galaxy type. \citet{Munoz-Mateos:09} have shown that dust attenuation for spiral galaxies has a large scatter and is nearly constant for mid- and late-type spirals. For early-type spirals, the attenuation decreases.
    
    \begin{figure}
        \includegraphics[width=8.8cm]{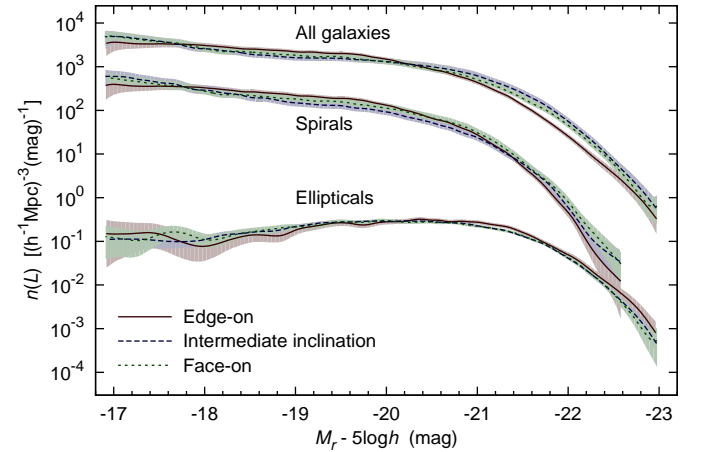}
        \caption{Attenuation-corrected LF for edge-on, face-on, and intermediate inclination angle galaxies. LFs have shown for spirals, ellipticals, and all galaxies. The LFs of spirals and all galaxies are shifted up by three and four units in logarithmic scale, for better separation of these three types. The filled areas show the 95\% confidence regions of the LF.}
        \label{fig:lf_incl2}
    \end{figure}

    Additionally, dust attenuation also depends on the inclination angle, as predicted by \citet{Tuffs:04} and confirmed also observationally by \citet{Driver:07}. Recently, dust attenuation in M\,31 was studied by \citet{Tempel:10}; the derived model allowed to investigate also the attenuation dependence on the inclination angle, as shown in Fig.~\ref{fig:nld_model}. At lower inclination angles (nearly face-on oriented galaxies) the total extinction is low, since the line-of-sight optical depth of the dust disc is low. Attenuation increases when moving toward higher inclination angles and reaches its maximum for nearly edge-on galaxies. For exactly edge-on galaxies, dust attenuation decreases again, as the thickness of the dust disc is generally much lower than the thickness of optical components. Figure~\ref{fig:nld_model} shows that the dust attenuation -- inclination angle relation depends also on the shape of the galaxy. For bulge dominated galaxies, the relation is flatter and attenuation is larger. For disc dominated galaxies, attenuation is lower and increases more rapidly while moving toward edge-on orientation.

    The inclination-dependent attenuation curves, used in this paper are shown in Fig.~\ref{fig:nld_model}. To classify galaxies into bulge-dominated or disc-dominated spirals, we use the parameter $f_{\mathrm{deV}}$: for lower values, the galaxy is disc-dominated and for larger values, the galaxy is bulge-dominated. We use a linear interpolation in $f_{\mathrm{deV}}$ when moving from disc-dominated galaxies to bulge-dominated galaxies. Since we do not know how the attenuation exactly depends on the inclination angle and galaxy colour, we have used only the general trends for correcting the LFs statistically. The calibration of the corrections has been carried out by comparing the LFs for galaxies at different inclination angles. For each galaxy subtype (colour) we modify the attenuation curves given in Fig.~\ref{fig:nld_model}, to minimise differences between the shapes of the LFs for different inclination angles. For that, we multiply the attenuation curve by a factor $x$ that depends on the galaxy rest-frame \hbox{$u$-$r$} colour. For red spiral galaxies (with \hbox{$u$-$r>2.2$}), $x=0.5$ provided the best fit. This is expected, because red galaxies tend to contain less dust. For galaxies with \hbox{$u$-$r<1.8$} $x=1.0$. For the intermediate galaxies, the factor $x$ changes linearly with the \hbox{$u$-$r$} colour.

    Figure~\ref{fig:lf_incl} shows the LFs for the observed and attenuation-corrected luminosities for the edge-on and the face-on samples of galaxies. For the edge-on sample $\cos{\theta}<0.2$, for the face-on sample $\cos{\theta>0.8}$. It is well seen that the observed LFs are quite different for the face-on and edge-on galaxies, while the attenuation-corrected LFs are quite similar. Figure~\ref{fig:lf_incl2} shows the attenuation-corrected luminosity functions for the spirals and the ellipticals and for all galaxies together at different inclination angles: edge-on, face-on, and intermediate inclination angles. The intermediate inclination angle sample is defined by $0.5<\cos{\theta}<0.6$. We see that the LF is nearly inclination independent. For elliptical galaxies, the LF is absolutely inclination independent. For the edge-on spirals and for all edge-on galaxies together, small differences are still noticeable.

    \section{Results} 

    For all of the LFs presented in this section, the Schechter and double-power-law parameters are given in Table~\ref{table:lf}.
    
    \begin{figure}
        \includegraphics[width=8.8cm]{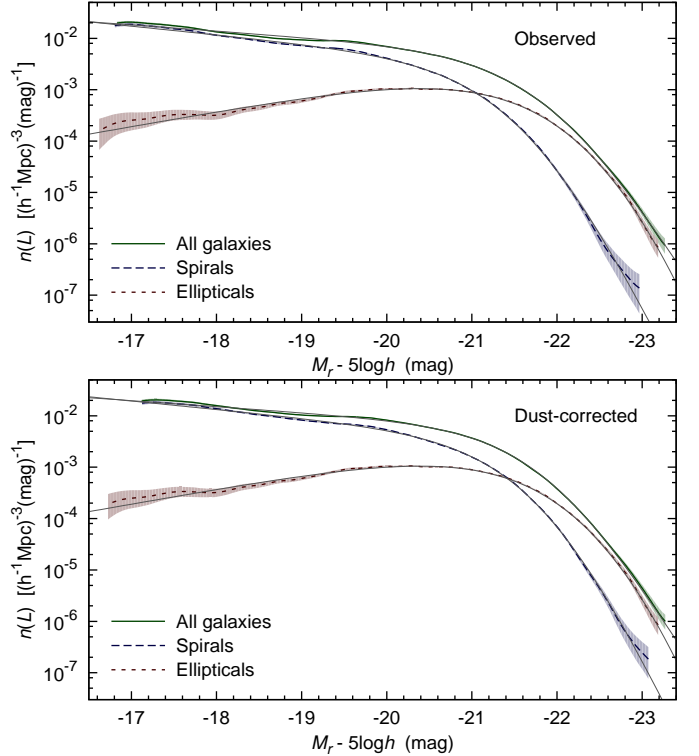}
        \caption{LFs for all galaxies, spirals, and ellipticals. \emph{Upper panel}: observed LFs. \emph{Lower panel}: attenuation-corrected LFs. The filled areas show the 95\% confidence regions of the LF. Light grey lines are analytical double-power-law functions.}
        \label{fig:lf_alltype}
    \end{figure}

    \begin{figure*}
        \includegraphics[width=180mm]{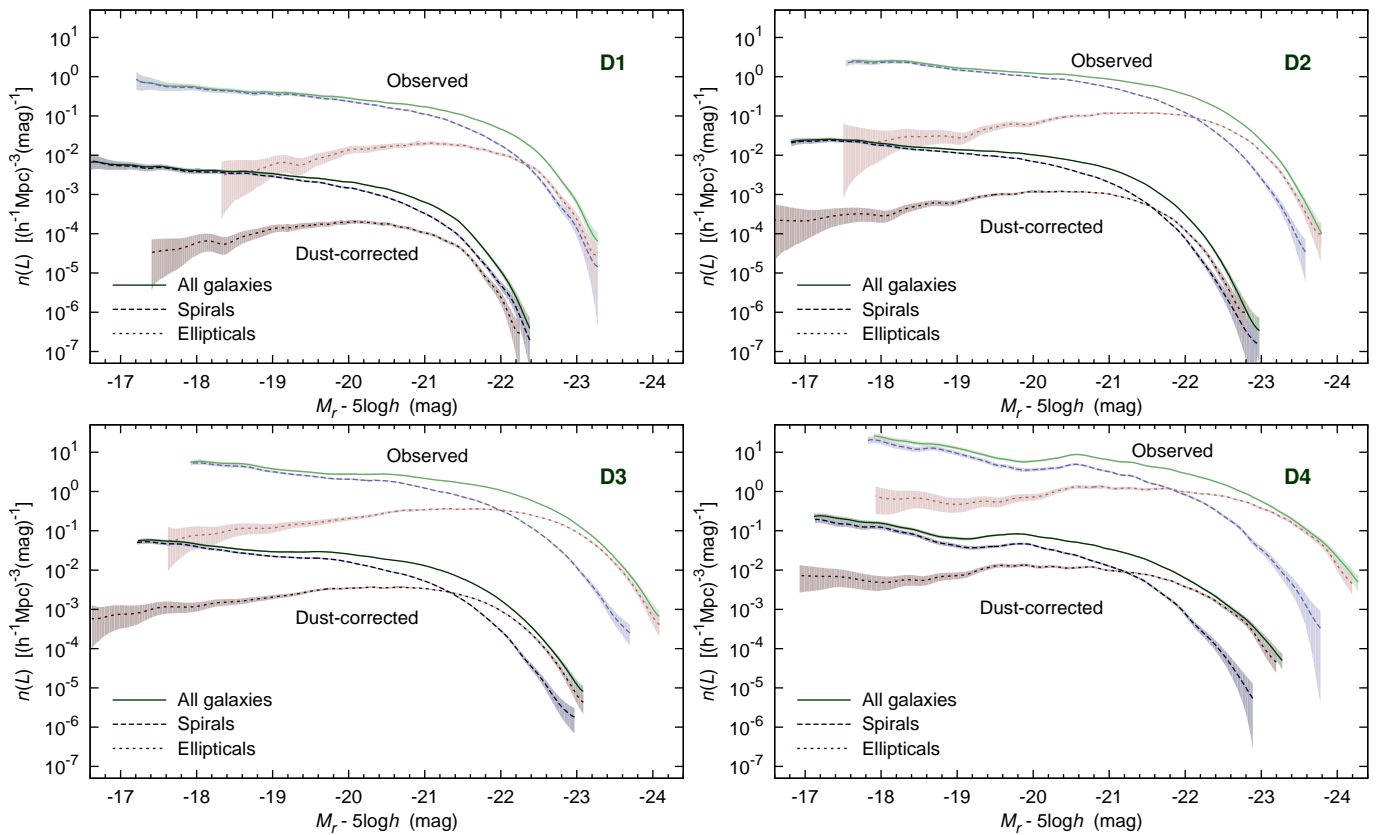}
        \caption{LFs for different types of galaxies in different environments: \emph{upper left} panel shows the least dense environment (D1) and \emph{bottom right} panel shows the most dense environment (D4). Green solid lines show the LFs for all galaxies; blue dashed lines show the LFs for spiral galaxies; red dotted lines show the LFs for elliptical galaxies. The upper LFs in each panel are the observed LFs that are shifted up two units in logarithmic scale and right by one mag. The lower LFs in each panel are the attenuation-corrected LFs. Filled areas show the 95\% confidence regions. The panels D1 to D4 show different global environmental regions: D1 are void regions and D4 are supercluster regions. The LFs have been normalised to the volume of each sample.}
        \label{fig:lf_densities}
    \end{figure*}
    
    \begin{figure}
        \includegraphics[width=8.8cm]{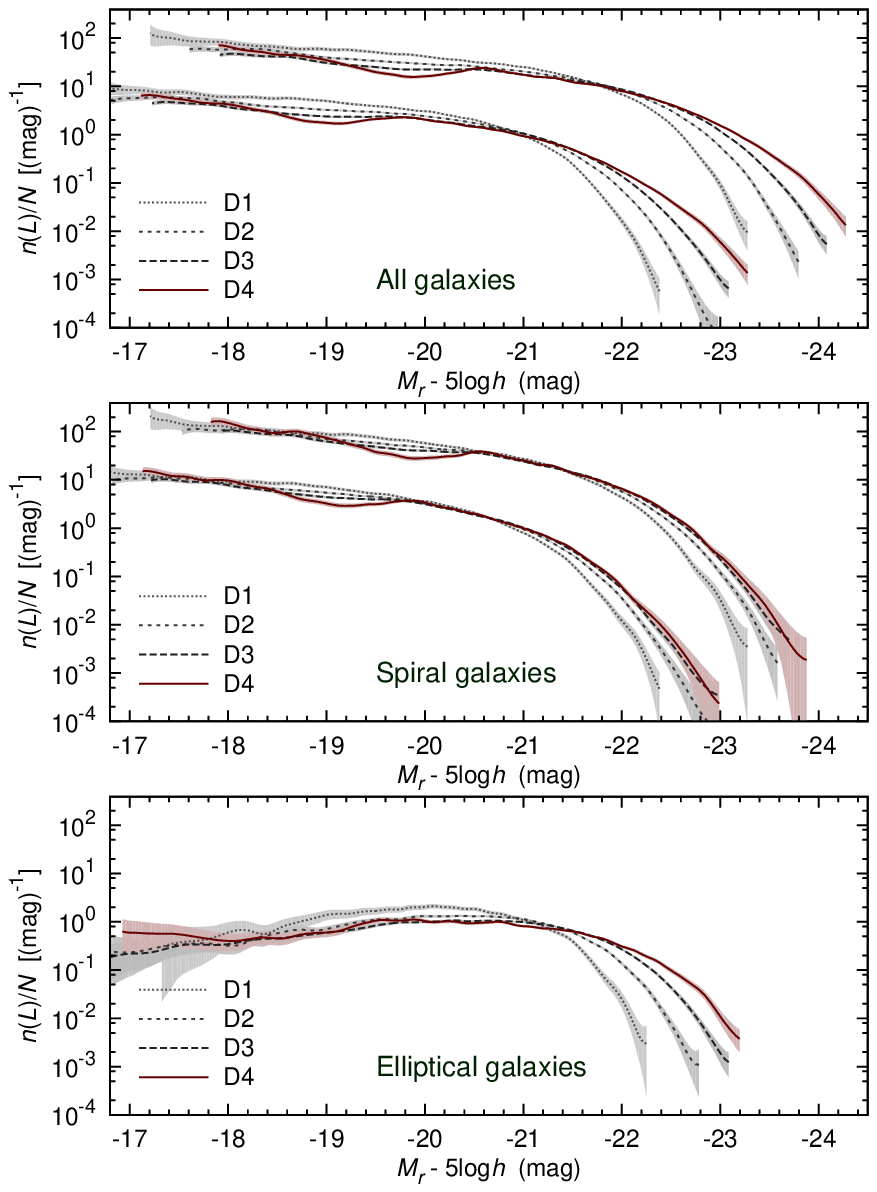}
        \caption{LFs for different types of galaxies in different environments. The upper LFs in each panel are the observed LFs that are shifted up two units in logarithmic scale and right by one mag. The lower LFs in each panel are the attenuation-corrected LFs. For elliptical galaxies (\emph{bottom panel}), only the observed LFs are shown. Dotted lines stand for the least dense (void) environments; red solid lines stand for the most dense (supercluster) environments. The filled areas show the 95\% confidence regions of the LFs.}
        \label{fig:lf_types_dens}
    \end{figure}

    \subsection{Attenuation-corrected luminosity function} 

    Figure~\ref{fig:lf_alltype} shows the LFs for all galaxies, for spirals, and for ellipticals in the attenuation-corrected (lower panel) and uncorrected (upper panel) case. At the bright end, most of the galaxies are ellipticals, and at the faint end, most of the galaxies are spirals, as found in many previous studies. Correction for dust attenuation increases the brightness of spiral galaxies; the shift is especially noticeable at the bright-end of the LF. However, even with the attenuation correction applied, the brightest spiral galaxies are still less luminous than the brightest elliptical galaxies by about 0.5\,mag. Table~\ref{table:lf} gives the analytical fits to the LFs for the double-power law as well for the Schechter function. In Fig.~\ref{fig:lf_alltype} the double-power-law fits are shown as solid lines for illustration.

    \subsection{Luminosity functions in different environments} 

    Figure~\ref{fig:lf_densities} shows the LFs in different environments: D1 to D4, where D1 is the least dense (void) environment and D4 is the most dense (supercluster) environment. In each panel, lower and darker lines are the attenuation-corrected LFs and upper, dimmer lines are the observed LFs. The most notable trend is that while moving from the lower global densities toward the higher ones, elliptical galaxies start to dominate the bright-end of the LF. In the least dense environments, ellipticals and spirals are equally abundant at the bright-end of the LF; in the densest environments, the brightest galaxies are mostly ellipticals. At the faint-end of the LF, while moving from the low density regions to the high density regions, the difference between ellipticals and spirals decreases; in denser environments, the fraction of elliptical galaxies increases. In general, differences between the attenuation-corrected and uncorrected cases are similar for all environments.

    A notable feature in the LF of ellipticals in the densest environments is a local minimum near  $M_r =$ -18\,mag. A seemingly similar feature is present for spirals at $M_r\approx -19$\,mag. However, these minima are of different origin: in the case of spirals, the small dip becomes visible because of an interplay with the bump at $M_r\approx -19.8$\,mag, mainly caused by a selection effect. The number of galaxies in the most dense environments is relatively small in the SDSS sample volume and the presence or absence of a rich supercluster (the most dense environments) at a given distance interval may leave a notable feature in the LFs because of the apparent luminosity limits of the survey. At $M_r\approx -19$\,mag, the distance interval between 150 and 200\,h$^{-1}$Mpc determines most of the LF (see Fig.~\ref{fig:dist_vs_mag}). However, no rich superclusters are found in this region.

    In Fig.~\ref{fig:lf_types_dens} the LFs for spiral galaxies, for elliptical galaxies, and for all galaxies together are shown for different environments. For elliptical galaxies, the bright-end of the LF moves toward higher luminosities when moving toward higher densities. This means that bright elliptical galaxies are residing mostly in high density environments, e.g., in the cores of galaxy clusters.

    Interestingly, the LF for spiral galaxies is almost independent of environment. The faint-end of the LF of the most dense environments is slightly different, but the number of galaxies in this region is also small and the dip of the LF may be caused purely by selection effects in the SDSS, as mentioned above. The bright-end of the LF of the least dense environments is also slightly different from that in other environments, because generally, very bright galaxies are absent from the low density environments \citep{Tempel:09}.

    Comparing the LFs of spiral galaxies and elliptical galaxies, the LFs for spirals slightly increase at the faint end in all environments, while the LFs of ellipticals have a maximum at about $M_r\approx -20$ and decrease toward lower luminosities. The LF for all galaxies is a combination of the spiral LF and the elliptical LF: it is determined by ellipticals at the bright end and by spirals at the faint end.

    Figure~\ref{fig:lf_dencol} shows the LFs for red and blue ellipticals and spirals separately in different global environments. In general, the faint-end of the LF is mostly built up by bluer galaxies and the bright end includes mostly redder galaxies; this behaviour is the same for spirals and ellipticals.

    From Fig.~\ref{fig:lf_dencol} we also see that the increase of the number of bright ellipticals in dense environments is mostly caused by red ellipticals. As mentioned above, the LF of spirals is independent of the global environmental density (see Fig.~\ref{fig:lf_types_dens}). However, small changes with environment are seen at the faint-end of the LF. Figure~\ref{fig:lf_dencol} shows that this change is much smaller for the red and blue spirals separately and is increased by the interplay of the differences in the LF shapes of these subpopulations. Once again, these differences concern only the densest environments and are thus a subject for selection effects.

    \section{Discussion} 

    \subsection{Interpretation of the results} 

    Our analysis of the LFs of galaxies of different morphology in different global environments shows that the global environment has played an important role in galaxy evolution.

    An interesting result of the present study is the finding that the LF of spiral galaxies is almost independent of the global environment, especially when looking at the red and blue spirals separately. However, one has to keep in mind that only the normalised distributions are similar; the actual number densities of spiral galaxies are different. On the contrary, for elliptical galaxies, the global environment plays an important role: brighter elliptical galaxies are located mostly in denser environments. In the case of ellipticals, the environment is more important for the red galaxies than for the blue ones.

   The results suggest that the evolution of spiral galaxies is slightly different for different types (colours) of spirals, but for a fixed-type spiral galaxy, the evolution is independent of the global environmental density. Thus the formation history of spiral galaxies in various global environments has to be similar. This result seems to be contradicting with the general $\Lambda$CDM cosmology: galaxy luminosity should be linked to the mass of the parent dark matter halo, the distribution of which depends on the environment. Besides, according to \citet{Delgado-Serrano:10}, approximately half of the spirals were already in place 6~Gyr ago and another half formed in mergers of irregular galaxies. This suggests that the minor mergers and quiescent star formation are the dominant factors that determine the formation of spiral galaxies. A possible interpretation of our results may lie in the fragility of spiral galaxies: they form and survive only in specific conditions (e.g. the preservation of the gas, the absence of major mergers) which are typical of low density regions, but to some extent can be present also in high density regions. Some haloes may remain intact and host a spiral galaxy regardless of the global environment, while most of the potential hosts of spiral galaxies end up hosting ellipticals as demonstrated by the semi-analytical models \citep{Benson:10a}.

    The derived LF of elliptical galaxies can be reconciled with the hierarchical galaxy formation through mergers. The denser the environment, the brighter galaxies there should reside because of the increased merger rate. The difference between the LFs of elliptical galaxies in different environments is more notable for red galaxies, in accordance with their supposed merger origin. This interpretation agrees well also with the picture of hierarchical formation of galaxies: for blue galaxies, the evolution is more quiescent and major mergers are not so important; for red ellipticals, merging is the dominant factor of galaxy evolution. Since blue ellipticals are most likely S0-s or late-type ellipticals, they have still some gas available for star formation and therefore the evolution of blue ellipticals is closer to the evolution of spiral galaxies -- the global environment is less important.

    \begin{figure*}
          \includegraphics[width=18cm]{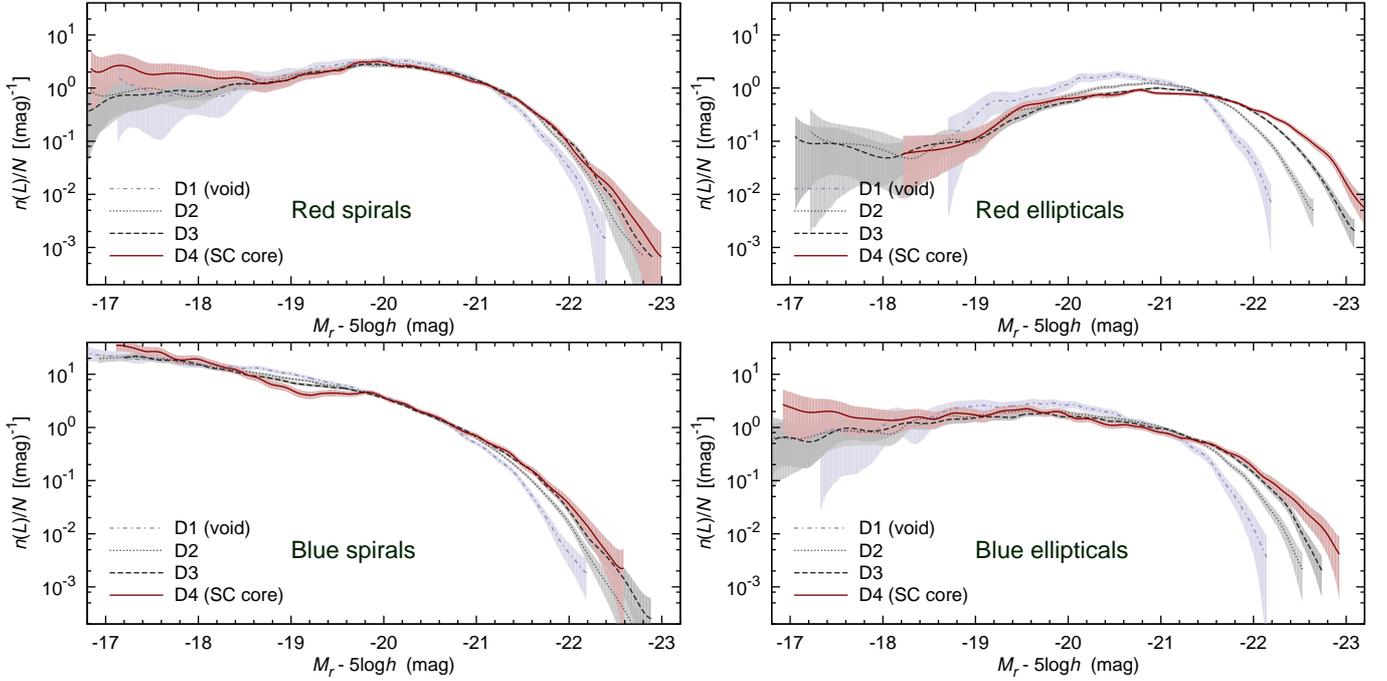}
          \caption{Attenuation-corrected LFs for red (\emph{upper panels}) and blue (\emph{lower panels}); for spirals (\emph{left panels}) and ellipticals (\emph{right panels}), in different global environments. D1 is the least dense environment (voids); D4 is the most dense environment (superclusters). The filled areas show the 95\% confidence regions of the LFs.}
          \label{fig:lf_dencol}
      \end{figure*}


    \subsection{Comparison with earlier work} 

    Influence of the global environment on the LF has been previously studied by \citet{Hoyle:05} using the SDSS data and by \citet{Croton:05} using the 2dFGRS data. These works have shown that galaxies in the void environment are primarily of late-type. In our results, this effect becomes especially pronounced after applying the absorption corrections: Fig.~\ref{fig:lf_densities} shows that in void regions, spiral galaxies dominate over ellipticals at all luminosities.
    
    The LFs derived by \citet{Hoyle:05} for a considerably smaller sample of galaxies are quite similar in different environments, except for the highest density environment, where the faint-end slope is shallower. \citet{Croton:05} have found that the faint-end of the LF depends weakly on environment. In general, our analysis confirms this result, except for the highest density environment, where an excess of faint galaxies compared to other environments is found (noticeable for red spirals and for blue ellipticals in Fig.~\ref{fig:lf_dencol}). This excess has been detected also by \citet{Xia:06}. In deep surveys of the Hubble Space Telescope, an excess of faint red galaxies has been found in the field environments: \citet{Salimbeni:08} have seen such trend in the GOODS dataset and \citet{Drory:09} in the COSMOS field. Thus the excess of faint red galaxies appears in all environments, most strongly in dense cluster regions.

    \citet{Phleps:07} have studied the global environment beyond the redshift 2. Using three different fields with different global environments, they show that for blue galaxies, the environment plays a smaller role than for red galaxies. Their results are in agreement with our findings for the relatively nearby region.

    Many previous works were concentrated only on the local environment. For example, \citet{Yang:09} studied the LF for the central and satellite galaxies in groups, using the SDSS data. They found that in general, red galaxies are the central galaxies and blue galaxies are satellite galaxies; however, they found that for very low masses, the number of red central galaxies increases. They speculate that these galaxies are located close to large haloes so that their star formation is truncated by the large scale environment. Our results also show that the faint-end of the LF increases when moving toward very high density (see Fig.~\ref{fig:lf_types_dens}). When splitting our galaxies into the red and blue samples (see Fig.~\ref{fig:lf_dencol}), the increase of the LF at the faint end is noticeable for every subsample (see Table~\ref{table:lf}), indicating a universal trend. In our case, the increase is also noticeable for blue galaxies. However, a direct comparison with the results of \citet{Yang:09} is difficult, because the environmental densities have been estimated differently.

    \citet{Zandivarez:06} shows that the local environment (galaxy group mass) is an important factor in galaxy evolution. They show that the faint-end slope is practically constant for the blue cloud galaxies, while for the red sequence galaxies, the faint end is steeper for more massive systems. Their results can be interpreted in terms of galaxy mergers as the main driving force behind galaxy evolution in groups. Using local environmental densities, we get similar results: for ellipticals (that are dominantly red), the faint-end slope is changing with density, and for spirals (that are dominantly blue), the faint-end slope is practically constant. We shall discuss the group environment in more details in a forthcoming study, using the group catalogue from \citet{Tago:10}.
    
    \begin{figure}
        \includegraphics[width=8.8cm]{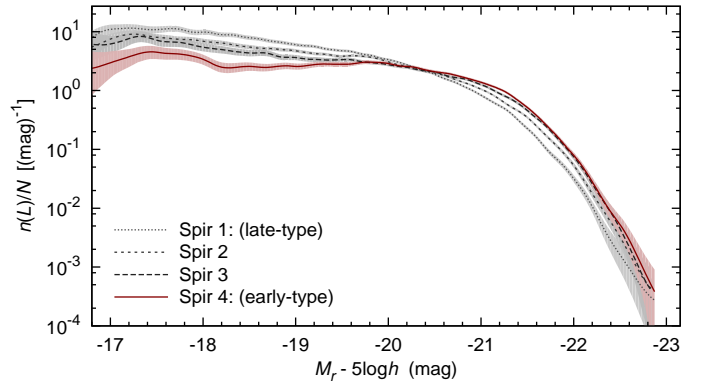}
        \caption{The attenuation-corrected LF for spiral galaxies for various Hubble types.The Hubble type indicator is the value of $f_{\mathrm{deV}}$: low value -- late-type; high value -- early-type. Errors are omitted for clarity. The filled areas show the 95\% confidence regions of the LF.}
        \label{fig:lf_frac2}
    \end{figure}

    In general, the characteristic magnitudes for ellipticals are brighter than for spirals and the faint-end slopes are steeper for spirals. \citet{Devereux:09} used the $K$-band luminosity to derive the LF for different Hubble types; classification was performed visually. The average shape of the LFs of ellipticals and spirals is generally in agreement with our results. In addition, they found that the faint-end slope is steeper for late-type spirals than for early-type (S0) spirals and that the characteristic luminosity is larger for early-type galaxies. In Fig.~\ref{fig:lf_frac2} we use the value $f_{\mathrm{deV}}$ to separate spiral galaxies into early and late type. This figure shows a similar trend as pointed out by \citet{Devereux:09}, but in our case, the differences are smaller. The qualitative results presented in Fig.~\ref{fig:lf_frac2} remain the same, when using the classification by \citet{Huertas-Company:11}.

    \section{Conclusions} 

    We have used the SDSS data to derive the LF of spiral galaxies, elliptical galaxies, and all galaxies together in various environments. We have taken special care to correct the galaxy luminosities for the intrinsic attenuation. The principal results of our study are the following:
    \begin{itemize}
        \item The LF of elliptical galaxies depends strongly on the environment; this suggests that global environmental density is an important driving force (via merging history) of elliptical galaxy formation. Density environment is more important for red elliptical galaxies than for blue elliptical galaxies.
        \item The evolution of spiral galaxies (the LF of spiral galaxies) is almost independent of environment, especially for blue and red spirals separately, showing that spiral galaxy formation has to be similar regardless of the surrounding global density.
        \item The highest global density regions (superclusters) are significantly different from other regions, as stated also by \citet{Tempel:09}. Notably more faint galaxies are found in high density regions than in other environments.
        \item The brightest galaxies are absent from the void regions. After correcting for the intrinsic absorption, spiral galaxies dominate the LF of void regions at every luminosity.
        \item The faint-end of the LF is determined by spiral galaxies and the bright end by elliptical galaxies. The faint end includes mostly blue galaxies and the bright end mostly red galaxies.
        \item Detailed studies of LFs require galaxy luminosities to be corrected for the intrinsic absorption by dust. Dust absorption affects mostly the bright-end of the LF. For the full LF, including all galaxies, the characteristic luminosity increases after attenuation correction. The faint-end slope of the LF is practically independent on dust attenuation.
    \end{itemize}

    A comparison of these results with predictions of numerical simulations and/or semianalytical models would provide stringent constraints on the driving factors of the formation and evolution of galaxies in dark matter haloes.  


    \begin{acknowledgements}
        We thank the referee for useful comments that helped to improve the paper. We acknowledge the financial support from the Estonian Science Foundation {grants 7115, 7146, 7765, 8005} and the Estonian Ministry for Education and Science research project SF0060067s08. This work has also been supported by Astrophysikalisches Institut Potsdam (using DFG-grant Mu~1020/15-1), where part of this study was performed. We thank P.~Einasto and I.~Enkvist for helping to classify galaxies visually, and E.~Tago for fruitful collaboration. E.T. thanks also P.~Tenjes for helpful discussion and suggestions. All the figures have been made with the gnuplot plotting utility.

        We are pleased to thank the SDSS Team for the publicly available data releases. Funding for the SDSS and SDSS-II has been provided by the Alfred P. Sloan Foundation, the Participating Institutions, the National Science Foundation, the U.S. Department of Energy, the National Aeronautics and Space Administration, the Japanese Monbukagakusho, the Max Planck Society, and the Higher Education Funding Council for England. The SDSS Web Site is \texttt{http://www.sdss.org/}.

        The SDSS is managed by the Astrophysical Research Consortium for the Participating Institutions. The Participating Institutions are the American Museum of Natural History, Astrophysical Institute Potsdam, University of Basel, University of Cambridge, Case Western Reserve University, University of Chicago, Drexel University, Fermilab, the Institute for Advanced Study, the Japan Participation Group, Johns Hopkins University, the Joint Institute for Nuclear Astrophysics, the Kavli Institute for Particle Astrophysics and Cosmology, the Korean Scientist Group, the Chinese Academy of Sciences (LAMOST), Los Alamos National Laboratory, the Max-Planck-Institute for Astronomy (MPIA), the Max-Planck-Institute for Astrophysics (MPA), New Mexico State University, Ohio State University, University of Pittsburgh, University of Portsmouth, Princeton University, the United States Naval Observatory, and the University of Washington.
    \end{acknowledgements}
    
    \begin{table*}
        \caption{The parameters for the Schechter and double-power law LFs.}
        \label{table:lf}
        \centering
            \begin{tabular}{lcc|cccc}
                \hline\hline
                & \multicolumn{2}{c|}{Schechter} & \multicolumn{4}{c}{Double-power law} \\
                Sample & $\alpha$ & $M^*$ & $\alpha$ & $\gamma$ & $\delta$ & $M^*$ \\
                \hline
                All galaxies (observed)& $-1.12\pm 0.01$ & $-20.71\pm 0.01$ &
                $-1.30\pm 0.01$ & $1.73\pm 0.05$       & $-7.78\pm 0.36$ & $-21.98\pm 0.06$ \\
                Spiral galaxies (observed)   & $-1.18\pm 0.01$ & $-20.17\pm 0.01$ &
                $-1.42\pm 0.01$ & $1.74\pm 0.05$       & $-9.52\pm 0.52$  & $-21.69\pm 0.07$ \\
                Elliptical galaxies (observed)       & $-0.19\pm 0.01$ & $-20.59\pm 0.01$ &
                $-0.26\pm 0.02$  & $1.24\pm 0.03$      & $-14.64\pm 1.27$  & $-22.96\pm 0.14$ \\
                \hline
                All galaxies (D1, observed)       & $-0.98\pm 0.02$ & $-20.04\pm 0.01$ &
                $-1.36\pm 0.01$  & $2.00\pm 0.05$      & $-12.70\pm 1.13$  & $-21.81\pm 0.08$ \\
                All galaxies (D2, observed)       & $-0.98\pm 0.02$ & $-20.42\pm 0.01$ &
                $-1.32\pm 0.01$  & $1.79\pm 0.03$      & $-14.18\pm 1.21$  & $-22.36\pm 0.08$ \\
                All galaxies (D3, observed)       & $-1.01\pm 0.01$ & $-20.74\pm 0.01$ &
                $-1.24\pm 0.01$  & $1.64\pm 0.06$      & $-10.49\pm 1.11$  & $-22.42\pm 0.12$ \\
                All galaxies (D4, observed)       & $-1.24\pm 0.02$ & $-21.21\pm 0.03$ &
                $-1.18\pm 0.05$  & $1.32\pm 0.18$      & $-8.47\pm 1.98$  & $-22.58\pm 0.50$ \\
                Spiral galaxies (D1, observed)       & $-1.10\pm 0.01$ & $-19.81\pm 0.01$ &
                $-1.35\pm 0.01$  & $1.49\pm 0.03$      & $-18.43\pm 2.00$  & $-22.17\pm 0.16$ \\
                Spiral galaxies (D2, observed)       & $-1.14\pm 0.01$ & $-20.09\pm 0.01$ &
                $-1.40\pm 0.01$  & $1.63\pm 0.04$      & $-12.01\pm 0.86$  & $-21.91\pm 0.08$ \\
                Spiral galaxies (D3, observed)       & $-1.11\pm 0.01$ & $-20.20\pm 0.01$ &
                $-1.40\pm 0.02$  & $1.74\pm 0.07$      & $-9.82\pm 0.88$  & $-21.80\pm 0.11$ \\
                Spiral galaxies (D4, observed)       & $-1.19\pm 0.03$ & $-20.36\pm 0.03$ &
                $-1.49\pm 0.04$  & $2.17\pm 0.25$      & $-8.02\pm 1.26$  & $-21.56\pm 0.18$ \\
                Elliptical galaxies (D1, observed)       & \phantom{$-$}$0.41\pm 0.02$ & $-19.74\pm 0.01$ &
                \phantom{$-$}$0.39\pm 0.15$  & $1.28\pm 0.15$      & $-15.00\pm 2.00$  & $-22.10\pm 0.50$ \\
                Elliptical galaxies (D2, observed)       & \phantom{$-$}$0.19\pm 0.01$ & $-20.19\pm 0.01$ &
                $-0.14\pm 0.03$  & $1.45\pm 0.05$      & $-18.00\pm 2.00$  & $-22.68\pm 0.21$ \\
                Elliptical galaxies (D3, observed)       & $-0.08\pm 0.01$ & $-20.60\pm 0.01$ &
                $-0.11\pm 0.04$  & $1.18\pm 0.06$      & $-21.00\pm 2.00$  & $-23.42\pm 0.47$ \\
                Elliptical galaxies (D4, observed)       & $-0.45\pm 0.01$ & $-21.01\pm 0.01$ &
                $-0.42\pm 0.05$  & $1.05\pm 0.09$      & $-21.00\pm 2.00$  & $-24.02\pm 0.50$ \\
                \hline
                All galaxies (dust-corrected)& $-1.05\pm 0.01$ & $-20.66\pm 0.01$ &
                $-1.29\pm 0.01$ & $1.91\pm 0.05$   & $-7.69\pm 0.28$ & $-21.91\pm 0.05$ \\
                Spiral galaxies (dust-corrected)    & $-1.15\pm 0.01$ & $-20.32\pm 0.01$ &
                $-1.41\pm 0.01$ & $1.79\pm 0.05$   & $-9.88\pm 0.54$ & $-21.88\pm 0.06$   \\
                \hline
                All galaxies (D1, dust-corrected)       & $-0.98\pm 0.02$ & $-20.05\pm 0.02$ &
                $-1.32\pm 0.01$  & $2.06\pm 0.05$      & $-12.60\pm 0.91$  & $-21.83\pm 0.07$ \\
                All galaxies (D2, dust-corrected)       & $-0.96\pm 0.02$ & $-20.44\pm 0.01$ &
                $-1.28\pm 0.01$  & $1.83\pm 0.03$      & $-15.11\pm 1.14$  & $-22.39\pm 0.07$ \\
                All galaxies (D3, dust-corrected)       & $-0.98\pm 0.01$ & $-20.73\pm 0.01$ &
                $-1.23\pm 0.01$  & $1.72\pm 0.06$      & $-10.92\pm 1.15$  & $-22.41\pm 0.12$ \\
                All galaxies (D4, dust-corrected)       & $-1.23\pm 0.02$ & $-21.20\pm 0.02$ &
                $-1.21\pm 0.05$  & $1.49\pm 0.17$      & $-8.26\pm 1.99$  & $-22.43\pm 0.35$ \\
                Spiral galaxies (D1, dust-corrected)       & $-1.04\pm 0.01$ & $-19.98\pm 0.01$ &
                $-1.33\pm 0.01$  & $1.59\pm 0.04$      & $-14.60\pm 1.55$  & $-22.06\pm 0.12$ \\
                Spiral galaxies (D2, dust-corrected)       & $-1.08\pm 0.01$ & $-20.23\pm 0.01$ &
                $-1.28\pm 0.02$  & $1.54\pm 0.05$      & $-12.00\pm 1.12$  & $-22.09\pm 0.12$ \\
                Spiral galaxies (D3, dust-corrected)       & $-1.11\pm 0.01$ & $-20.40\pm 0.01$ &
                $-1.33\pm 0.02$  & $1.77\pm 0.08$      & $-9.00\pm 0.67$  & $-21.84\pm 0.09$ \\
                Spiral galaxies (D4, dust-corrected)       & $-1.19\pm 0.03$ & $-20.52\pm 0.03$ &
                $-1.29\pm 0.06$  & $1.82\pm 0.23$      & $-9.02\pm 1.95$  & $-21.85\pm 0.26$ \\
                \hline
                Red spirals (D1, dust-corrected)       & \phantom{$-$}$0.46\pm 0.01$ & $-19.62\pm 0.01$ &
                \phantom{$-$}$0.37\pm 0.05$  & $1.22\pm 0.05$      & $-20.02\pm 2.00$  & $-22.39\pm 0.27$ \\
                Red spirals (D2, dust-corrected)       & \phantom{$-$}$0.14\pm 0.01$ & $-19.90\pm 0.01$ &
                \phantom{$-$}$0.04\pm 0.05$  & $1.25\pm 0.05$      & $-16.59\pm 2.00$  & $-22.41\pm 0.22$ \\
                Red spirals (D3, dust-corrected)       & $-0.02\pm 0.02$ & $-20.04\pm 0.01$ &
                \phantom{$-$}$0.03\pm 0.05$  & $1.18\pm 0.06$      & $-14.40\pm 2.00$  & $-22.41\pm 0.23$ \\
                Red spirals (D4, dust-corrected)       & $-0.16\pm 0.03$ & $-20.09\pm 0.02$ &
                $-0.34\pm 0.08$  & $1.58\pm 0.15$      & $-8.35\pm 1.03$  & $-21.65\pm 0.16$ \\
                Blue spirals (D1, dust-corrected)       & $-1.26\pm 0.01$ & $-19.84\pm 0.01$ &
                $-1.28\pm 0.03$  & $1.12\pm 0.05$      & $-20.81\pm 2.00$  & $-22.68\pm 0.58$ \\
                Blue spirals (D2, dust-corrected)       & $-1.38\pm 0.01$ & $-20.12\pm 0.01$ &
                $-1.41\pm 0.02$  & $1.13\pm 0.04$      & $-20.10\pm 2.00$  & $-22.90\pm 0.38$ \\
                Blue spirals (D3, dust-corrected)       & $-1.47\pm 0.01$ & $-20.34\pm 0.01$ &
                $-1.55\pm 0.03$  & $1.27\pm 0.08$      & $-12.55\pm 2.00$  & $-22.39\pm 0.34$ \\
                Blue spirals (D4, dust-corrected)       & $-1.56\pm 0.03$ & $-20.52\pm 0.03$ &
                $-1.86\pm 0.04$  & $2.14\pm 0.03$      & $-7.90\pm 1.84$  & $-21.67\pm 0.28$ \\
                Red ellipticals (D1, dust-corrected)       & \phantom{$-$}$1.70\pm 0.04$ & $-19.52\pm 0.01$ &
                \phantom{$-$}$0.96\pm 0.14$  & $1.57\pm 0.16$      & $-20.00\pm 2.00$  & $-22.20\pm 0.46$ \\
                Red ellipticals (D2, dust-corrected)       & \phantom{$-$}$1.31\pm 0.01$ & $-19.96\pm 0.01$ &
                \phantom{$-$}$1.04\pm 0.07$  & $1.32\pm 0.06$      & $-20.00\pm 2.00$  & $-22.71\pm 0.27$ \\
                Red ellipticals (D3, dust-corrected)       & \phantom{$-$}$0.77\pm 0.01$ & $-20.41\pm 0.01$ &
                \phantom{$-$}$0.75\pm 0.05$  & $1.17\pm 0.05$      & $-20.00\pm 2.00$  & $-23.23\pm 0.29$ \\
                Red ellipticals (D4, dust-corrected)       & \phantom{$-$}$0.27\pm 0.02$ & $-20.81\pm 0.01$ &
                \phantom{$-$}$0.87\pm 0.20$  & $0.81\pm 0.13$      & $-20.00\pm 2.00$  & $-24.07\pm 0.50$ \\
                Blue ellipticals (D1, dust-corrected)       & $-0.03\pm 0.02$ & $-19.74\pm 0.01$ &
                $-0.14\pm 0.09$  & $1.22\pm 0.12$      & $-20.00\pm 2.00$  & $-22.49\pm 0.78$ \\
                Blue ellipticals (D2, dust-corrected)       & $-0.30\pm 0.01$ & $-20.19\pm 0.01$ &
                $-0.52\pm 0.04$  & $1.34\pm 0.08$      & $-20.00\pm 2.00$  & $-22.82\pm 0.49$ \\
                Blue ellipticals (D3, dust-corrected)       & $-0.48\pm 0.01$ & $-20.48\pm 0.01$ &
                $-0.63\pm 0.03$  & $1.27\pm 0.07$      & $-20.00\pm 2.00$  & $-23.15\pm 0.57$ \\
                Blue ellipticals (D4, dust-corrected)       & $-0.88\pm 0.02$ & $-20.86\pm 0.02$ &
                $-0.61\pm 0.12$  & $0.85\pm 0.16$      & $-20.00\pm 2.00$  & $-24.16\pm 0.50$ \\
                \hline
            \end{tabular}
        \tablefoot{$M^*$ is in units of mag$-5\log(h)$.}
    \end{table*}

    \begin{appendix}
        \section{Confidence intervals for luminosity functions}
        \label{app:1}

        The standard $V_{\mathrm{max}}^{-1}$ weighting
        procedure finds the differential luminosity function $n(L)\mathrm{d}L$ (the
        expectation of the number density of galaxies of the luminosity $L$)
        as follows
        \begin{equation}
          n(L) {\mathrm d}L = \sum_i\frac{\mathbf{I}_{(L,L+\mathrm{d}L)}(L_i)}
          {V_{\mathrm{max}}(L_i)},
          \label{eq:appVmax}
        \end{equation}
        where $\mathrm{d}L$ is the luminosity bin width, $\mathbf{I}_A(x)$ is
        the indicator function that selects the galaxies belonging to a
        particular luminosity bin, $V_{\mathrm{max}}(L)$ is the maximum volume
        where a galaxy of a luminosity $L$ can be observed in the present
        survey, and the sum is over all galaxies of the survey.

        This approach gives us the binned density histogram that depends both on the bin widths and the locations of the bin edges; a better way is to use kernel smoothing \citep[see, e.g.][]{wand95}, where the density is represented by a sum of kernels centred at the data points:
        \begin{equation}
          n(L) = \frac1h\sum_i\frac1{V_{\mathrm{max}}(L_i)}K\left(\frac{L-L_i}{h}\right).
          \label{eq:appK}
        \end{equation}
        The kernels $K(x)$ are distributions ($K(x)>0$, $\int K(x)\,dx=1$) of zero mean and of a typical width $h$. The width $h$ is an analogue of the bin width, but there are no bin edges to worry about.

        As the luminosity function is rapidly changing with luminosity, especially at the bright-end of the LF, the bin widths should vary. This is most easy to implement by adaptive kernel estimation of the LF -- instead of \ref{eq:appK} we write
        \begin{equation}
          n(L) =\sum_i\frac1{V_{\mathrm{max}}(L_i)}\frac1{h_i}K\left(\frac{L-L_i}{h_i}\right),
          \label{eq:appKad}
        \end{equation}
        where the kernel widths depend on the data, $h_i=h(L_i)$.
        
        The choice of the kernel widths is a matter of ongoing study, but recommendations are available \citep[see, e.g.][]{silverman86}. The kernel widths are known to depend on the density $f(x)$ itself, with $h\sim f(x)^{-1/5}$ for densities similar to  normal distribution. This choice requests a pilot estimate for the density that can be found using a constant width kernel. 

        We used the magnitude scale for our luminosity function (all kernel widths are in magnitudes), and the $B_3$ box spline kernel:
        \begin{eqnarray}
        B_3(x)&=&\left(|x-2|^3-4|x-1|^3+6|x|^3-\right.\nonumber\\
                &&\left.-4|x+1|^3+|x+2|^3\right)/12.
        \end{eqnarray}
        This kernel is well suited for estimating densities -- it is compact, differing from zero only in the interval $x\in[-2,2]$, and it conserves mass: $\sum_i B_3(x-i)=1$ for any $x$.

        For the pilot estimate, we used a wide kernel with the scale $h=0.5$\,mag. For the adaptive kernel widths, we adopted $h=0.05$\,mag (the typical SDSS rms magnitude error) as the minimal width (for the maximum pilot density) and rescaled it by the $h\sim f_{\mathrm{pilot}}(x)^{-1/5}$ law. The luminosity function drops sharply at the bright end, leading to very wide kernels; we restricted the kernel width by $2h=0.5$\,mag.

        If we choose the kernel width this way, we minimise the mean integrated standard error (MISE) of the density. While that is certainly a useful quantity, we are also interested in the ``error bars'', pointwise confidence intervals for the density. These can be obtained by smoothed bootstrap \citep{silverman87, davison97, fiorio04}. Here the data points for the bootstrap realisations are chosen, as usual, randomly from the observed data with replacement, but they have an additional smoothing component:
        \begin{equation}
          L^\star_i =L_j+h\varepsilon_j,   
          \label{eq:appsmb}
        \end{equation}
        where $\varepsilon$ is a random variable of the density $K(x)$.
        
        We generated 10000 bootstrap realisations, using the adaptive kernel widths as for the true luminosity function estimate. We show the centred 95\% confidence regions in our figures.

    \end{appendix}

\end{document}